\documentclass{elsarticle}
\usepackage{xcolor}
\usepackage{colortbl}
\usepackage{graphicx}
\usepackage{amsmath}

\begin{document}

\title{The superharmonic instability and wave breaking in Whitham equations}

\author{John D.~Carter\corref{cor1}}
\ead{carterj1@seattleu.edu}
\address{Mathematics Department, Seattle University, USA}

\author{Marc Francius}
\ead{marc.francius@mio.osupytheas.fr}
\address{Université de Toulon, Aix Marseille Univ, CNRS, IRD, MIO, Toulon, France}

\author{Christian Kharif}
\ead{christian.kharif@centrale-marseille.fr}
\address{Aix-Marseille Université, CNRS, Centrale Marseille, IRPHE, UMR 7342, 13384, Marseille, France.}

\author{Henrik Kalisch}
\ead{Henrik.Kalisch@uib.no}
\address{Department of Mathematics, PO Box 7800, 5020 Bergen,Norway}

\author{Malek Abid}
\ead{malek.abid@univ-amu.fr}
\address{Aix-Marseille Université, CNRS, Centrale Marseille, IRPHE, UMR 7342, 13384, Marseille, France.}

\date{\today}

\begin{abstract}

    The Whitham equation is a model for the evolution of surface waves on shallow water that combines the unidirectional linear dispersion relation of the Euler equations with a weakly nonlinear approximation based on the KdV equation.  We show that large-amplitude, periodic, traveling-wave solutions to the Whitham equation and its higher-order generalization, the cubic Whitham equation, are unstable with respect to the superharmonic instability (i.e.~a perturbation with the same period as the solution).  The threshold between superharmonic stability and instability occurs at the maxima of the Hamiltonian and $\mathcal{L}_2$-norm.  We examine the onset of wave breaking in traveling-wave solutions subject to the modulational and superharmonic instabilities.  
    
    We present new instability results for the Euler equations in finite depth and compare them with the Whitham results.  We show that the Whitham equation more accurately approximates the wave steepness threshold for the superharmonic instability of the Euler equations than does the cubic Whitham equation.  However, the cubic Whitham equation more accurately approximates the wave steepness threshold for the modulational instability of the Euler equations than does the Whitham equation.

\end{abstract}

\maketitle

\section{Introduction}

White-capping through spilling and micro breaking is a ubiquitous feature of ocean waves, is a key component of air-sea interaction, and is known to be a significant factor in the kinetic and thermal energy budgets of the ocean.  While much of white-capping is driven by surface winds and wave group behavior, in many cases, early theoretical studies of wave hydrodynamics relevant to breaking considered the simplest approach (ideal fluid, irrotational flow and negligible wind effects) and focussed on a number of hydrodynamic instabilities of two-dimensional uniform wave trains in deep water.  Now it is known that wave breaking can be induced by an instability in the crest of a steep wave, a so-called crest instability that corresponds to a form of the superharmonic instability of a progressive gravity wave. This instability has been studied in a great many works, usually within the framework of the fully nonlinear potential Euler equations (also known as the ``water-wave problem''). Our aim is to show that the crest instability is captured by simplified models for water waves, assumed to be weakly nonlinear but fully dispersive.  In the process, we also provide some new instability results for the full water-wave problem.

Longuet-Higgins \cite{longuet1978instabilities} was the first to find that very steep Stokes wave trains are linearly unstable with respect to perturbations of the same wavelength and phase-locked to the basic wave. He conjectured the existence of an exchange of stability for the wave whose phase velocity is a maximum. Later on, Tanaka \cite{tanaka1983stability}, using a more accurate approach, found that the exchange of stability occurs at the maximum of the energy and not at the maximum of phase velocity.  Using Zakharov's Hamiltonian formulation, Saffman \cite{saffman1985superharmonic} proved analytically that an exchange of stability occurs when the wave energy is an extremum as a function of the wave height.  Furthermore, he confirmed the non-existence of superharmonic bifurcation predicted by Tanaka \cite{tanaka1985stability}.  Recently, Sato \& Yamada \cite{sato2019superharmonic} revisited Saffman's theorem and showed that the exchange of stability occurs when the energy is stationary as a function of the wave velocity. Zufiria \& Saffman \cite{zufiria1986superharmonic} extended Saffman's theorem to the case of finite depth.  Kataoka \cite{Kataoka} revisited the work of Zufiria \& Saffman analytically and numerically and found that the superharmonic instability threshold for periodic waves on fluids of finite depth occurs at the maximum of the Hamiltonian.  Note that Tanaka \cite{tanaka1986stability} found that very steep solitary waves are subject to crest instability of superharmonic type, too.  Within the framework of the potential Euler equations, Francius \& Kharif~\cite{FK1} suggested and provided preliminary numerical results for a dimensionless depth of $d=2$ on the existence of the occurrence of the superharmonic instability at the maximum of the energy.  Tanaka {\em et al.} \cite{tanaka1987instability} used a boundary integral method to show that the nonlinear evolution of the crest instability leads to the overturning of the solitary wave.  Longuet-Higgins \& Dommermuth \cite{longuet1997crest} showed numerically that the nonlinear development of the crest instability of periodic gravity waves produces the overturning of the wave crest depending on the sign of the unstable perturbation.

Due to the computational complexity of the Euler equations, it has long been of interest to find a simpler model equation that allows smooth periodic and solitary waves, but also the existence of highest waves with singularities at the crest as observed with the Euler equations. In this vein, Whitham~\cite{Whitham} was the first to propose a simplified nonlocal water-wave model, the so-called Whitham equation, by combining the unidirectional linear dispersion relation of the Euler equations with a weakly nonlinear approximation based on the KdV equation for improving the description of the dynamics of weakly nonlinear long-waves.  Much later on, Ehrnstr\"om \& Kalisch\cite{EK} demonstrated rigorously the existence of traveling periodic wave solutions of the Whitham equation, and Ehrnstr\"om \& Wahl\'en~\cite{WhithamCusp} proved that the highest traveling-wave solutions with maximal wave height are cusped. 

On one hand, Hur \& Johnson~\cite{HurJohnson2015} proved that small-amplitude traveling-wave solutions of the Whitham equation are stable with respect to the modulational instability if $k<1.146$ and are unstable with respect to the modulational instability if $k>1.146$ where $k$ is a dimensionless wavenumber of the solution, equivalent to the dimensionless depth.  Sanford {\emph{et al.}}~\cite{Sanford2014} and Carter \& Rozman~\cite{STWhitham} numerically studied the stability of traveling-wave solutions to the Whitham equation.  They corroborated the Hur \& Johnson $k=1.146$ threshold and showed that large-amplitude traveling-wave solutions are unstable regardless of their wavelength.  Adding a higher-order term, Carter {\emph{et al.}}~\cite{CVWhitham} studied the stability of solutions to the cubic Whitham equation and found results qualitatively similar to those in the Whitham equation.  Using the method described in Binswanger {\emph{et al.}}~\cite{Binswanger} corroborates the Hur \& Johnson threshold and shows that small-amplitude traveling-wave solutions to the cubic Whitham equation are unstable with respect to the MI when $k>1.252$.  These values should be compared with the well-known critical value $k_c=1.363$ for small amplitude gravity waves in the Euler equations. On the other hand, up to the best of our knowledge, no information is available on the linear stability of periodic traveling-wave solutions subject to superharmonic disturbances within the framework of the Whitham equations. Nonetheless, we remark that, very recently, Bronski {\emph{et al.}}~\cite{BronskiHurWester} demonstrated analytically and numerically that periodic traveling waves of certain regularized long-wave models are linearly unstable to superharmonic perturbations. Examples analyzed by these authors include the regularized Boussinesq, Benney-Luke, and Benjamin-Bona-Mahony equations. The purpose of the present paper is therefore to analyze the superharmonic instability of traveling wave solutions within the context of either the Whitham and cubic Whitham equations.

The remainder of this paper is organized as follows.  Section \ref{SectionWhitham} contains a brief introduction to the Whitham and cubic Whitham equations.  Section \ref{SectionStability} contains a numerical study of the linear stability for the traveling-wave solutions of these equations.  Comparisons of the results from the Whitham and cubic Whitham equations with those from the Euler equations are also presented in this section.  Section \ref{SectionBreaking} contains a numerical examination of the nonlinear stability and the onset of wave breaking in traveling-wave solutions perturbed by modulational and superharmonic instabilities.  Section \ref{SectionSummary} contains a summary of our results.

\section{The Whitham and cubic Whitham equations}
\label{SectionWhitham}

In dimensionless variables, the Whitham equation is given by
\begin{equation}
    u_t+\mathcal{K}*u_x+\frac{3}{2}uu_x=0,
    \label{Whitham}
\end{equation}
where $\mathcal{K}$ is the kernel of the convolution operator defined in terms of its Fourier transform
\begin{equation}
    \hat{\mathcal{K}}(\kappa)=\sqrt{\frac{\tanh(\kappa)}{\kappa}}.
\end{equation}
where $\kappa$ is the wavenumber in Fourier space.  Here $u=u(x,t)$ represents the dimensionless surface displacement.  The Whitham equation can be converted to dimensional form via the transformation
\begin{equation}
    x\rightarrow h_0x,~~~~~t\rightarrow \sqrt{\frac{h_0}{g}}~t,~~~~~u\rightarrow h_0u,
\end{equation}
where $h_0$ is the dimensional undisturbed fluid depth and $g$ represents the acceleration due to gravity.  

Whitham \cite{Whithambook} conjectured that (\ref{Whitham}) would be more suitable for describing the evolution of water waves since it does not have the long-wavelength restriction inherent in models such as the KdV and Boussinesq equations.  Recent work has shown that the Whitham equation and some of its generalizations are able to describe surface waves more accurately than comparable long-wave models~\cite{moldabayev2015whitham,Trillo,WhithamComp,emerald2021rigorous}.
    
Kharif \& Abid~\cite{Kharif2018} extended equation (13.131) of Whitham~\cite{Whithambook} for potential flows to flows of constant vorticity.  Expanding this new generalized Whitham equation to second order in amplitude and setting the vorticity to zero gives the cubic Whitham equation
\begin{equation}
    u_t+\mathcal{K}*u_x+\frac{3}{2}uu_x - \frac{3}{8} u^2u_x=0.
    \label{CWhitham}
\end{equation}

These two evolution equations possess a Hamiltonian structure.  They can be written as
\begin{equation}
    u_t=J\frac{\delta\mathcal{H}}{\delta u},
\end{equation}
where $J=-\partial_x$ represents a skew-symmetric linear operator and $\frac{\delta\mathcal{H}}{\delta u}$ is the variational derivative of the Hamiltonian functional.  Equation (\ref{Whitham}) has Hamiltonian
\begin{equation}
    \mathcal{H}_W=\frac{1}{2}\int_{-L/2}^{L/2}\left( u\mathcal{K}*u+\frac{1}{2}u^3 \right)dx,
    \label{Ham3}
\end{equation}
and equation (\ref{CWhitham}) has Hamiltonian
\begin{equation}
    \mathcal{H}_{cW}=\frac{1}{2}\int_{-L/2}^{L/2}\left( u\mathcal{K}*u+\frac{1}{2}u^3-\frac{1}{16}u^4\right)dx,
    \label{Ham4}
\end{equation}
where $L=L_0/h_0$ and $L_0$ is the dimensional spatial period of the solution.  Note that with this scaling, the dimensionless wavenumber $k=2\pi/L$ coincides with the dimensionless undisturbed depth $d=h_0 k_0$.  As is well known from the Hamiltonian representation of evolution equations, the invariance of these equations under translations along the $t$-axis implies that both equations preserve their Hamiltonians in $t$.  In addition, these equations have two other classical conserved quantities
\begin{equation}
    \mathcal{M}=\int_{-L/2}^{L/2}u~dx,\hspace*{1cm}\mathcal{L}_2=\int_{-L/2}^{L/2}u^2~dx,
\end{equation}
which correspond to the mass and the impulse of the solution respectively. Note that the invariance in $t$ of the impulse, namely the $\mathcal{L}_2$-norm of the solution, is due to the invariance of these Hamiltonian systems under translations along the $x$-axis.

\section{Periodic traveling waves}
\label{SectionStability}

\subsection{Steady periodic traveling waves}
\label{SectionStability_1}
\begin{figure}
    \begin{center}
        \includegraphics[width=12cm]{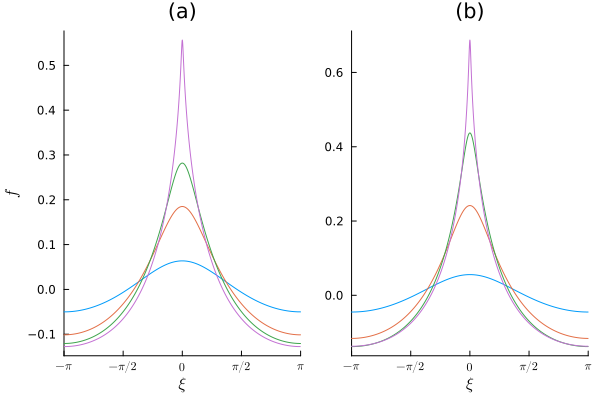}
        \caption{Plots of $2\pi$-periodic traveling-wave solutions to (a) the Whitham equation and (b) the cubic Whitham equation.}
        \label{SolnsPlot}
    \end{center}
\end{figure}

We computed periodic traveling-wave solutions of the form $u(x,t)=f(x-ct)=f(\xi)$ where $f$ is a smooth function and $c$ is a real constant using the branch-following method described in Ehrnstr\"om \& Kalisch~\cite{ehrnstrom2013global} and Carter {\emph{et al.}}~\cite{CVWhitham}  We only considered solutions with zero mean since they are the most physically relevant.  Plots of $2\pi$-periodic solutions to the Whitham and cubic Whitham equations are included in Figure \ref{SolnsPlot}.  The tallest solutions shown are close in wave height to the solutions with maximal height.  The values of the wave speed, $c$; the wave height, defined as the vertical distance between the crest and the trough, $H$; the wave steepness, $s=H/L$; the Hamiltonian, $\mathcal{H}$; and the $\mathcal{L}_2$-norm, $\mathcal{L}_2$; for these solutions are included in Table \ref{SolnsPlotsTable}.

\begin{table}
    \begin{center}
      \begin{tabular}{c|ccccc}
        Eqn, color & $c$ & $H$ & $s$ & $\mathcal{H}$ & $\mathcal{L}_2$ \\
        \hline
        W, blue & 0.878 & 0.114 & 0.018 & 0.004 & 0.010 \\
        W, orange & 0.903 & 0.286 & 0.046 &0.025 & 0.057 \\
        W, green & 0.928 & 0.403 & 0.064 & 0.044 & 0.098 \\
        W, magenta & 0.976 & 0.684 & 0.109 & 0.069 & 0.150 \\
        \arrayrulecolor{gray}\hline
        \arrayrulecolor{black}
        cW, blue & 0.876 & 0.101 & 0.016 & 0.003 & 0.008 \\
        cW, orange & 0.914 & 0.358 & 0.057 & 0.037 & 0.083 \\
        cW, green & 0.959 & 0.575 & 0.092 & 0.074 & 0.161 \\
        cW, magenta & 0.988 & 0.824 & 0.131 &0.089 & 0.193 \\
      \end{tabular}
      \caption{Parameter values for the solutions plotted in Figure \ref{SolnsPlot}.  The first column lists the equation and the color of the solution curve.  The parameters $c$, $H$, $s$, $\mathcal{H}$, and $\mathcal{L}_2$ represent the wave speed, wave height, wave steepness, Hamiltonian, and $\mathcal{L}_2$-norm, respectively.  The $\mathcal{H}$ column lists values for $\mathcal{H}_W$ for the Whitham equation and values of $\mathcal{H}_{cW}$ for the cubic Whitham equation.}
      \label{SolnsPlotsTable}
    \end{center}
  \end{table}

\begin{figure}
    \begin{center}
        \includegraphics[width=12cm]{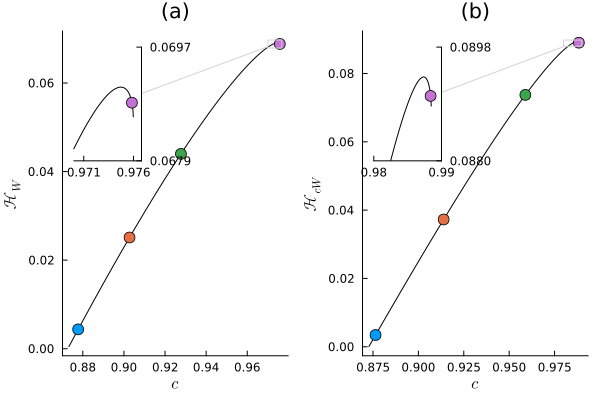}
        \caption{Plots of the Hamiltonians versus $c$ for the branches of $2\pi$-periodic solutions of (a) the Whitham equation and (b) the cubic Whitham equation.  The inset plots show zooms of the regions near the local maxima.}
        \label{HamPlot}
    \end{center}
\end{figure}

\begin{figure}
    \begin{center}
        \includegraphics[width=12cm]{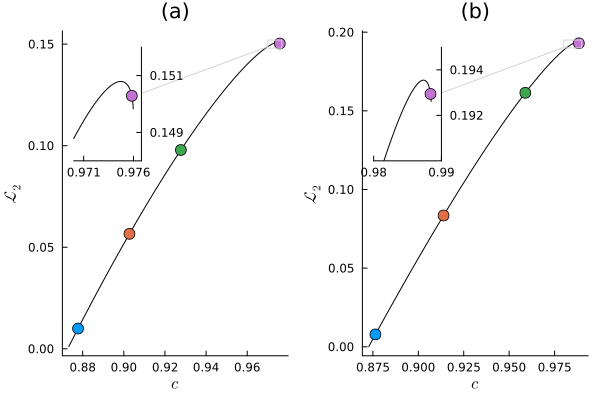}
        \caption{Plots of the $\mathcal{L}_2$-norm versus $c$ for the branches of $2\pi$-periodic solutions of (a) the Whitham equation and (b) the cubic Whitham equation.  The inset plots show zooms of the regions near the local maxima.}
        \label{L2Plot}
    \end{center}
\end{figure}

Figure \ref{HamPlot} contains plots of the Hamiltonians $\mathcal{H}_W$ and $\mathcal{H}_{cW}$ versus $c$ for the $2\pi$-periodic solution branches of the two equations.  The colored dots correspond to the solutions plotted in those colors in Figure \ref{SolnsPlot}.  For both equations, the Hamiltonians achieve local maxima at critical wave speeds, $c=c^*$.  Figure \ref{L2Plot} contains plots of the $\mathcal{L}_2$-norm versus $c$ for the $2\pi$-periodic solution branches of both equations.  Note that the $\mathcal{L}_2$-norms also achieve local maxima at the same critical value $c=c^*$.  It is no coincidence that the extrema of the Hamiltonian and the $\mathcal{L}_2$-norm occur at the same critical value.  In fact, traveling waves correspond to critical points of an augmented Hamiltonian functional $\mathcal{H}_{aug}=\mathcal{H}(u)-c\mathcal{L}_2(u)+b\mathcal{M}(u)$ for some real $b$.  For solutions with zero mean (i.e.~$\mathcal{M}=0$), the corresponding Euler-Lagrange equation is
\begin{equation}
    \frac{\delta\mathcal{H}}{\delta u}-c\frac{\delta\mathcal{L}_2}{\delta u}=0.
\end{equation}

Using the notation $\mathcal{H}=\hat{H}(c,L)$ and $\mathcal{L}_2=\hat{L}_2(c,L)$ for the conserved quantities evaluated along the two-parameter branch of periodic traveling-wave solutions of the Whitham equation, then it follows that
\begin{equation}
    \frac{d\hat{H}}{dc}-c\frac{d\hat{L}_2}{dc}=0,
\end{equation}
for a fixed wavelength $L$.  Thus, the two quantities are stationary at the same value of $c$.  See Benjamin~\cite{BenjaminVariation} for details.

The values of the parameters corresponding to the maxima of the Hamiltonians (and $\mathcal{L}_2$-norms) are listed in Table \ref{ParameterStarTable} as starred values.  Due to the resolution required to resolve solutions near the (cusped) solutions with maximal wave height using a Fourier basis, we were unable to determine if the Hamiltonians continue to decrease monotonically after the local maxima or if local minima are achieved for some $c>c^*$.  We note that the plot of the Hamiltonian in the Euler equations on infinite depth case oscillates many times~\cite{SHInstabStokesWaves}.

\begin{table}
    \begin{center}
      \begin{tabular}{c|cccccc}
        Eqn, $L$ & $c^\dagger$ & $H^\dagger$ & $s^\dagger$ & $\mathcal{H}^\dagger$ & $\mathcal{L}_2^\dagger$ \\
        \hline
        W, $\pi$ & 0.694 & 0.000 & 0.000 & 0.000 & 0.000 \\
        W, $2\pi$ & 0.925 & 0.390 & 0.062 & 0.042 & 0.094 \\
        W, $3\pi$ & 1.049 & 0.600 & 0.064 & 0.099 & 0.201 \\
        W, $4\pi$ & 1.097 & 0.633 & 0.050 & 0.120 & 0.235 \\
        \arrayrulecolor{gray}\hline
        \arrayrulecolor{black}
        cW, $\pi$ & 0.694 & 0.000 & 0.000 & 0.000 & 0.000 \\
        cW, $2\pi$ & 0.953 & 0.544 & 0.087 & 0.069 & 0.152 \\
        cW, $3\pi$ & 1.069 & 0.753 & 0.080 & 0.134 & 0.269 \\
        cW, $4\pi$ & 1.119 & 0.805 & 0.064 & 0.166 & 0.321 \\
      \end{tabular}
      \caption{Values of the parameters at the onset of the modulational instability (daggered parameters).  The first column lists the equation and the period of the solution.  The parameters $c$, $H$, $s$, $\mathcal{H}$, and $\mathcal{L}_2$ represent the wave speed, wave height, wave steepness, Hamiltonian, and $\mathcal{L}_2$-norm, respectively.  The $\mathcal{H}^\dagger$ column lists values for $\mathcal{H}^\dagger_W$ and $\mathcal{H}^\dagger_{cW}$.}
      \label{ParameterDaggerTable}
    \end{center}
  \end{table}

\begin{table}
    \begin{center}
      \begin{tabular}{c|cccccc}
        Eqn, $L$ & $c^*$ & $H^*$ & $s^*$ & $\mathcal{H}^*$ & $\mathcal{L}_2^*$ \\
        \hline
        W, $\pi$ & 0.766 & 0.526 & 0.167 & 0.021 & 0.058\\
        W, $2\pi$ & 0.974 & 0.647 & 0.103 & 0.069 & 0.151\\
        W, $3\pi$ & 1.058 & 0.671 & 0.071 & 0.102 & 0.206 \\
        W, $4\pi$ & 1.100 & 0.669 & 0.053 & 0.121 & 0.236 \\
        \arrayrulecolor{gray}\hline
        \arrayrulecolor{black}
        cW, $\pi$ & 0.773 & 0.623 & 0.198 & 0.025 & 0.069\\
        cW, $2\pi$ & 0.987 & 0.789 & 0.126 & 0.089 & 0.194\\
        cW, $3\pi$ & 1.076 & 0.836 & 0.089 & 0.137 & 0.274\\
        cW, $4\pi$ & 1.122 & 0.850 & 0.068 & 0.167 & 0.322\\
      \end{tabular}
      \caption{Values of the parameters at the onset of the superharmonic instability (starred parameters).  The first column lists the equation and the period of the solution.  The parameters $c$, $H$, $s$, $\mathcal{H}$, and $\mathcal{L}_2$ represent the wave speed, wave height, wave steepness, Hamiltonian, and $\mathcal{L}_2$-norm, respectively.  The $\mathcal{H}^*$ column lists values for $\mathcal{H}^*_W$ and $\mathcal{H}^*_{cW}$.}
      \label{ParameterStarTable}
    \end{center}
  \end{table}  

\subsection{Linear stability analysis}
\label{SectionStability_2}
We consider perturbed solutions of the form
\begin{equation}
    u_{pert}(\xi,t)=u(\xi)+\epsilon u_1(\xi,t)+\mathcal{O}(\epsilon^2),
    \label{PertForm}
\end{equation}
where $u$ is a traveling-wave solution with period $L$, $\xi=x-ct$, $\epsilon$ is a small constant, and $u_1$ is the leading-order term of the perturbation.  Using the Fourier-Floquet-Hill method described in Deconinck \& Kutz~\cite{DK}, assume
\begin{equation}
    u_1(\xi,t)=\mbox{e}^{i\mu \xi}U(\xi)\mbox{e}^{\lambda t}+c.c.,
    \label{vForm}
\end{equation}
where $\mu\in [-\pi/L,\pi/L]$ is known as the Floquet parameter, $\lambda$ is a complex constant, $c.c.$~stands for complex conjugate, and $U(\xi)$ is a function with period $L$ and Fourier series
\begin{equation}
    U(\xi)=\sum_{j=-N}^{N}\hat{U}(j)\mbox{e}^{2\pi i j\xi/L}.
    \label{VForm}
\end{equation}
Here $2N+1$ is the total number of Fourier modes and the $\hat{U}(j)$ are the complex amplitudes of the Bloch function $U(\xi)$ associated with the perturbation.  If $\mu=0$, then the perturbation has the same $\xi$-period as the unperturbed solution.  If there exists a perturbation with $\mu=0$ and $\Re(\lambda)>0$, then the solution is said to be linearly unstable with respect to the superharmonic instability (SI).  If there exists a perturbation with $\mu$ close to zero, but nonzero, and $\Re(\lambda)>0$, then the perturbation has an $\xi$-period that is larger than the unperturbed solution and the solution is said to be linearly unstable with respect to the modulational instability (MI).  Modulational instabilities are sometimes referred to as subharmonic instabilities.  The solution is said to be linearly stable if there does not exist any $\mu$ or $U$ such that $\Re(\lambda)>0$.  

For a branch of solutions corresponding to a given wavenumber, as the wave height of the solutions increases, the stability spectra for both equations go through two bifurcations.  First, the solutions become unstable with respect to the MI.  The second bifurcation occurs when the solutions become unstable with respect to the SI.  This second bifurcation point occurs at the maximum of the Hamiltonian and $\mathcal{L}_2$-norm.  Table \ref{ParameterDaggerTable} contains the values of the parameters where the first bifurcation occurs for solutions of both equations with four different periods.  For the solutions with $L=\pi$, the first bifurcation occurs at $H=0$ because small-amplitude solutions to both equations with $k=2$ are known to be unstable with respect to the MI~\cite{HurJohnson2015,CVWhitham,Binswanger}.  Table \ref{ParameterStarTable} contains the values of the parameters where the second bifurcation occurs.  Note that as the period of the solution increases, the values of the parameters at the threshold for the MI approach the values of the parameters at the threshold for the SI.  Also note that in the solitary-wave limit ($L\rightarrow\infty$) there is SI, but not MI.

\begin{figure}
    \begin{center}
        \includegraphics[width=12cm]{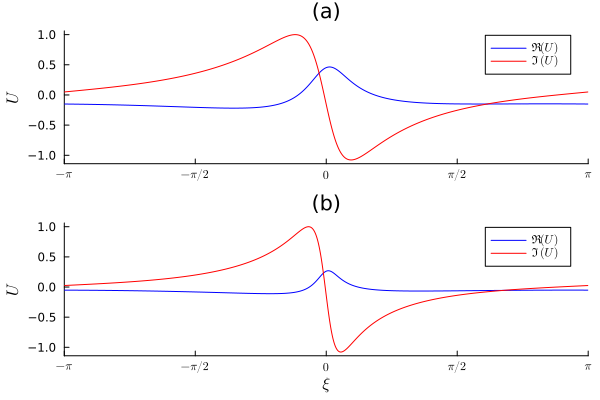}
        \caption{Plots of the real and imaginary parts of the periodic portion of the perturbations, $U(\xi)$, for the modulational instability when $\mu=0.1$ corresponding to the green solutions in Figure \ref{SolnsPlot} for the (a) Whitham and (b) cubic Whitham equations.}
        \label{MIPertPlots}
    \end{center}
\end{figure}

In Figure \ref{SolnsPlot}, the green curves are solutions with wave speeds slightly larger than $c^\dagger$.  These solutions are unstable with respect to the modulational instability, but not the superharmonic instability.  Figure \ref{MIPertPlots} includes plots of the real and imaginary parts of $U(\xi)$ corresponding to the unstable perturbations for these solutions when $\mu=0.1$.  In this case, the perturbations, $u_1(\xi,t)$ have a $\xi$-period of $20\pi$ and eigenvalues $\lambda=0.001503+0.01814i$ (Whitham) and $\lambda=0.002578+0.01694i$ (cubic Whitham).  The nonzero portion of the MI perturbations is centered at the peak of the unperturbed solution.  As the wave steepnesses of the solutions increase, the steepnesses of the MI perturbations also increase.

\begin{figure}
    \begin{center}
        \includegraphics[width=12cm]{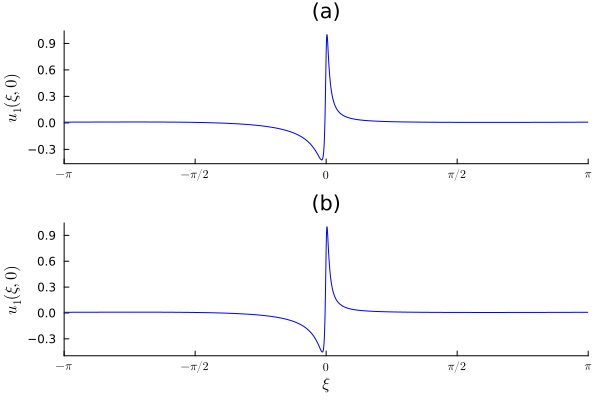}
        \caption{Plots of the superharmonic instability, $u_1(\xi,0)$, corresponding to the magenta solutions in Figure \ref{SolnsPlot} for the (a) Whitham and (b) cubic Whitham equations.}
        \label{SHPertPlots}
    \end{center}
\end{figure}

In Figure \ref{SolnsPlot}, the magenta curves are solutions with wave speeds slightly larger than $c^*$.  These solutions are unstable with respect to the superharmonic instability.  Figure \ref{SHPertPlots} includes plots of the superharmonic instability corresponding to these solutions.  These perturbations have the same period as the underlying solutions ($L=2\pi$) and have purely real eigenvalues of $\lambda=0.2517$ (Whitham) and $\lambda=0.2337$ (cubic Whitham).  This means that the perturbation is phase-locked with the basic wave.  Similarly to the MI, the nonzero portions of these perturbations are centered at the peaks of the solutions.  However, the superharmonic instabilities are significantly steeper than the modulational instabilities.

\subsection{Comparisons with results from the Euler equations}
\label{SectionWW}

In order to appreciate the differences between the Whitham and cubic Whitham equations and whether one of them constitutes a better approximation of the full Euler equations, we now compare the results of these three equations.

Borluk {\emph{et al.}}~\cite{Borluk} carried out a similar comparison with the KdV and Whitham equation, for different wavelengths ($L=\pi$, $2 \pi$ and $4\pi$) and wave heights. Comparing the bifurcation curves of each model, they showed that for the steady Whitham waves with $L=\pi$ compare more favorably to the Euler waves than the KdV waves. For larger wavelengths, $L\ge 2\pi$, the Whitham waves compare poorly to the Euler waves, the KdV waves with the largest wavelength appearing to be a better approximation of the Euler waves.

\begin{figure}[h!]
    \begin{center}
        \includegraphics[width=12cm]{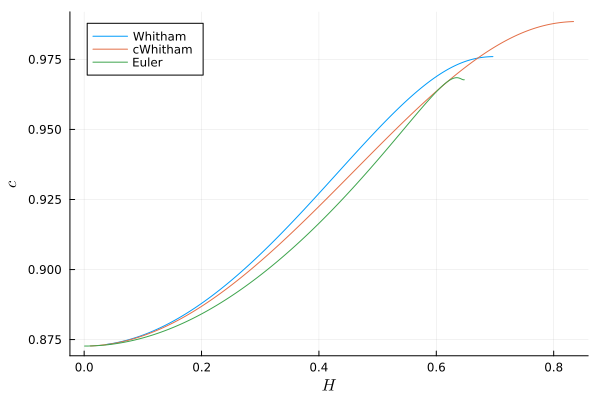}
        \caption{Plots of wave speed, $c$, versus wave height, $H$, for $2\pi$-periodic traveling-wave solutions of the Euler (with $d=1$), Whitham, and cubic Whitham equations.}
        \label{EulercHPlots}
    \end{center}
\end{figure}

Given these results, we have plotted in Figure \ref{EulercHPlots} the wave speed versus wave height curves for $2\pi$-periodic traveling-wave solutions of the Euler, Whitham, and cubic Whitham equations. Here, the steady waves of the Euler model were obtained with the numerical method proposed by Longuet-Higgins~\cite{longuet1978instabilities}. For any given undisturbed depth $d=2\pi/L$, this method enables the computation of the bifurcation branch with very high accuracy, up to wave heights close to the maximum value and certainly beyond the maximal value of the Hamiltonian (or maximal total energy).

The plots in Figure \ref{EulercHPlots} show that the results from the cubic Whitham equation are in better agreement with those from the Euler equations, in particular for the large wave heights.  However, the cubic Whitham equation admits solutions of significantly larger wave height and speed than do the Whitham and Euler equations.  According to the Euler equations, the limiting wave with period $L=2\pi$ has $s^{\prime}=0.1030$, $H^{\prime}=0.6473$ and $c^{\prime}=0.9690$. These estimates come from formulae (5.1) and (5.4) of Zhong \& Liao~\cite{Zhong}, which successfully obtained very accurate results (and especially the wave profiles) of the limiting Stokes’ waves in arbitrary water depth.

Considering the linear stability analysis of the Euler waves with $L=2\pi$ it is expected that for large enough wave height they become unstable to 1-D modulational instabilities, as evidenced by the Figure 3b in McLean~\cite{McLean}, which shows the bands of 1-D and 2-D instabilities for a wave with $H=0.580$ and $c=0.9588$.  Extending this work and using the numerical method described in Francius \& Kharif~\cite{FK2}, we found that the threshold steepness for the onset of the modulational instability in the Euler equations with $d=1$ is $s^\dagger=0.085$ ($c^\dagger=0.9477$).  These values should be compared with the threshold value $s^\dagger=0.062$ ($c^\dagger=0.925$) for the Whitham equation and $s^\dagger=0.087$ ($c^\dagger=0.953$) for the cubic Whitham equation, see Table \ref{ParameterDaggerTable}.  Hence the cubic Whitham result is closer to the exact value, which suggests that the addition of the cubic nonlinear in the Whitham equation provides an improvement for the $L=2\pi$ case.  The plots of Figure \ref{EulercHPlots} show that near this threshold value the bifurcation curve of the cubic Whitham equation is the closest one to that of the Euler equations.

\begin{figure}
    \begin{center}
        \includegraphics[width=12cm]{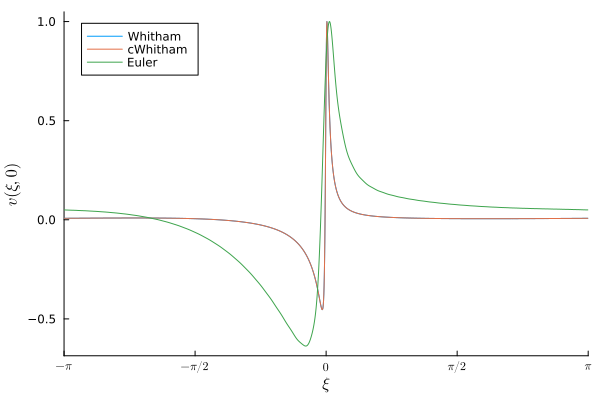}
        \caption{Plots comparing the superharmonic instability for the Whitham, cubic Whitham, and Euler equations for solutions with period $2\pi$.  Note that the Whitham and cWhitham plots lie essentially on top of one another.}
        \label{EulerSHInstabPlots}
    \end{center}
\end{figure}

For the Euler equations, the SI threshold occurs at the value $s^*=0.099$ ($c^*=0.968$).  These values should be compared with those from the Whitham ($s^*=0.103$, $c^*=0.974$) and cubic Whitham ($s^*=0.126$, $c^*=0.987$) equations, see Table \ref{ParameterStarTable}.  This establishes that the Whitham equation more accurately reproduces the SI threshold and corresponding speed of the Euler equations than does the cubic Whitham equation.  We compare the normalized profiles of the superharmonic instabilities corresponding to solutions with steepnesses slightly greater than the threshold value $s^*$.  Figure \ref{EulerSHInstabPlots} shows plots of the superharmonic instabilities for the $2\pi$-periodic solutions of the Whitham (from Figure \ref{SHPertPlots}(a) with $s=0.109$), cubic Whitham (from Figure \ref{SHPertPlots}(b) with $s=0.131$), and Euler equations ($s=0.100$).  The superharmonic instabilities for the Whitham and cubic Whitham equations are essentially the same and cannot be distinguished at the scale used in the figure.  However, the Euler instability is significantly less steep than the Whitham and cubic Whitham instabilities.  Finally we note that as the steepness of the solution increases, the steepnesses of the superharmonic instabilities also increase, as well as their growth rate (not shown here).
  
At first glance it may seem surprising that approximate models capture the crest instability.  However, the large-amplitude solutions to the Whitham and cWhitham equations are very steep and it is this steepness that triggers the superharmonic instability.  The cWhitham equation more accurately predicts the onset of the MI than does the Whitham equation while the Whitham equation more accurately predicts the onset of the SI.  Although there is qualitative agreement between the three models, there is not strong quantitative agreement.  

\section{Nonlinear instability and the onset of wave breaking}
\label{SectionBreaking}

In order to study the nonlinear stability of periodic traveling-wave solutions of the Whitham and cWhitham equations perturbed by modulational and superharmonic instabilities, we consider initial conditions of the form
\begin{equation}
    u_{0}(x,0)=u(x)+\epsilon u_1(x,0),
    \label{BreakingIC}
\end{equation}
where $u(x)$ is a periodic traveling-wave solution, $u_1(x,0)$ is a perturbation, and $\epsilon$ is a small real constant.  Initial conditions of this form were used in codes that time-evolve solutions of the Whitham and cWhitham equations.  The codes use a Fourier basis in space and the fourth-order operator-splitting technique introduced by Yoshida~\cite{yoshida} in time.  

\begin{figure}
    \begin{center}
        \includegraphics[width=12cm]{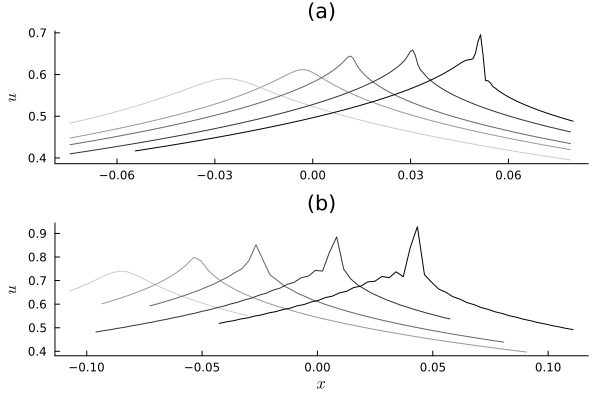}
        \caption{Plots demonstrating the evolution of the magenta solutions from Figure \ref{SolnsPlot} perturbed by the superharmonic instabilities from Figure \ref{SHPertPlots} with a positive value of $\epsilon$ near the onset of wave breaking for the (a) Whitham and (b) cWhitham equations.  The lighter curves evolve into the darker curves.  Note that the plots are zoomed in near the peaks of the solution and the solutions have been horizontally translated so that they can be compared side by side.}
        \label{SIBreakingPlot}
    \end{center}
\end{figure}

\begin{figure}
    \begin{center}
        \includegraphics[width=12cm]{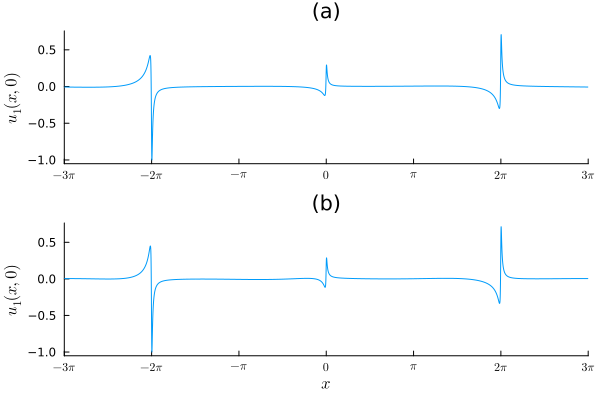}
        \caption{Plots of $u_1(x,0)$ used in Simulations \#3 and \#4 for the (a) Whitham and (b) cWhitham equations. This is the $\mu=1/3$ modulational instability corresponding to the magenta solutions in Figure \ref{SolnsPlot} .  Although these Whitham and cWhitham perturbations appear the same, they differ by as much as 10\% near the peaks.}
        \label{mu03PertPlots}
    \end{center}
\end{figure}

We ran a number of other simulations of special importance:
\begin{itemize}
    \item{{\textbf{Simulation \#1}}: Here we considered the superharmonic instability.  We used a single period of the magenta solutions in Figure \ref{SolnsPlot} as $u(x)$, the superharmonic instabilities shown in Figure \ref{SHPertPlots} as $u_1(x,0)$, and a positive value for $\epsilon$ in equation (\ref{BreakingIC}).  Initially, the perturbations grew exponentially with the rates predicted by linear theory ($\lambda=0.2517$ for Whitham and $\lambda=0.2337$ for cWhitham).  Then, nonlinear effects began to play a role and the solutions evolved towards breaking.  Figure \ref{SIBreakingPlot} contains plots of the perturbed solutions near the onset of breaking.}
    \item{{\textbf{Simulation \#2}}: This simulation was the same as Simulation \#1, except that we used a negative value for $\epsilon$.  In this case, the perturbations initially grew exponentially with the rates predicted by linear theory ($\lambda=0.2517$ for Whitham and $\lambda=0.2337$ for cWhitham).  However, when nonlinear effects began to play a role, the solutions did not evolve towards breaking.  Instead, they continued to evolve as perturbed traveling-wave solutions for a long period of time.}
    \item{{\textbf{Simulation \#3}}: In this case, we considered the $\mu=1/3$ modulational instability of solutions that are unstable with respect to both the modulational and superharmonic instabilities.  (Note that the $\mu=1/2$ modulational instability has the largest growth rate in this case.  However, we were able to monitor the growth of the $\mu=1/3$ instability because it was seeded as part of the initial condition.)  We used three periods of the magenta solutions from Figure \ref{SolnsPlot} as $u(x)$, one period of the $\mu=1/3$ modulational instabilities shown in Figure \ref{mu03PertPlots}, and a positive value for $\epsilon$ to construct the initial condition.  Initially, the perturbations grew with the rates predicted by linear theory ($\lambda=0.2571$ for Whitham and $\lambda=0.2428$ for cWhitham).  When nonlinear effects began to play a role, the rightmost and center peaks evolved towards breaking in a manner similar to what was observed in Simulation \#1.  The rightmost peaks tended towards breaking sooner than the center peaks.  This is consistent with the fact that the magnitudes of the perturbations near that peak were larger than at the center peaks.  The leftmost peaks evolved in a manner similar to what was observed in Simulation \#2 where no trend towards breaking was observed.  This is consistent with the fact that the perturbation essentially has a negative sign near that peak.}
    \item{{\textbf{Simulation \#4}}: This simulation is the same as Simulation \#3, except that we used a negative value for $\epsilon$.  Initially, the perturbations grew with the rates predicted by linear theory.  When nonlinear effects begin to play a role, the leftmost peaks evolved towards breaking similarly to what was observed in Simulation \#1.  The rightmost and center peaks evolved in a manner similar to Simulation \#2 where no trend towards breaking was observed.  This again emphasizes that the sign and/or phase of the perturbation plays an important role in the onset of wave breaking.}
    \item{{\textbf{Simulations \#5 and \#6}}: In this case, we considered the $\mu=1/3$ modulational instability of solutions that are not unstable with respect to the superharmonic instability, but are close in wave height to solutions that are unstable with respect to the superharmonic instability.  These solutions behaved similarly to the solutions in Simulation \#3 when $\epsilon$ was positive and similarly to the solutions in Simulation \#4 when $\epsilon$ is negative.  Simulation \#5 demonstrates that the MI instability may trigger the SI instability as shown by Longuet-Higgins \& Cokelet~\cite{LHCokelet} who solved the Euler equations numerically.  Nevertheless, Simulation \#6 suggests the existence of a threshold value of the wave steepness of the basic wave above which the MI instability may trigger the SI instability.}
\end{itemize}

Long-time simulations using initial conditions formed by perturbing the green solutions from Figure \ref{SolnsPlot} (solutions with just enough steepness to be unstable with respect to the MI, but not enough to be unstable with respect to the SI) with the MI perturbations shown in Figure \ref{MIPertPlots} did not tend towards breaking.  Initially the perturbations grew with the growth rates predicted by linear theory (0.001503 for the Whitham case and 0.002578 for the cWhitham case), then nonlinear effects occurred, and eventually the solution nearly recurred to their initial states.  

The Whitham breaking results are consistent with the recent work of McAllister {\emph{et al.}}~\cite{McAllisterBreaking}.  They showed that when the local surface slope surpasses $u_x=0.577$ in simulations of the Euler equations, then the solution would break.  Our simulations of the Whitham equation show that the perturbed green solution, which has a maximal local steepness of $u_x=0.1833$, does not tend towards breaking while the perturbed magenta solution, which has a maximal local steepness of $u_x=1.90$, tends towards breaking.  Determining a precise local steepness cutoff for breaking in the Whitham equation remains an open question.

\section{Summary}
\label{SectionSummary}

We have shown that periodic traveling-wave solutions with large enough amplitude to both the Whitham and cubic Whitham equations are unstable with respect to the superharmonic instability.  This means that large-amplitude traveling-wave solutions of these equations with period $L$ are unstable with respect to perturbations with period $L$.  We showed that the threshold between superharmonic stability and instability occurs at the maxima of the Hamiltonian and $\mathcal{L}_2$-norm.  This qualitatively aligns with the results from the Euler equations.

We presented new modulational and superharmonic instability wave steepness thresholds for the Euler equations in finite depth.  We showed that the cubic Whitham equation more accurately approximates the Euler modulational instability threshold than does the Whitham equation.  However, the Whitham equation more accurately approximates the Euler superharmonic instability threshold than does the cubic Whitham equation.  The cWhitham equation works better for waves of moderate steepness while the Whitham equation works better for waves or larger steepness.

We showed that the sign and/or phase of the perturbation determines whether a perturbed traveling-wave solution of the Whitham or cubic Whitham equation evolves towards breaking.  This qualitatively aligns with the results from the Euler equations.

These results show that the relatively simple Whitham and cubic Whitham equations possess some of the same properties of the Euler equations.  To our knowledge, these are the first superharmonic instability results for periodic solutions to approximate models of surface waves on finite depth.

\end{document}